\documentclass[manuscript,screen,nonacm]{acmart}

\makeatletter
\@twosidefalse
\@mparswitchfalse
\renewcommand\@titlefont{\fontsize{14}{16}\selectfont\bfseries} 
\makeatother

\AtBeginDocument{%
  }

\renewcommand\footnotetextcopyrightpermission[1]{} 
\begin{document}

\title[AnchoredAI]{AnchoredAI: Contextual Anchoring of AI Comments Improves Writer Agency and Ownership}

\author{Martin Lou}
\email{runkai02@student.ubc.ca}
\affiliation{%
  \institution{The University of British Columbia}
  \city{Vancouver}
  \country{Canada}
}

\author{Jackie Crowley}
\affiliation{%
  \institution{The University of British Columbia}
  \city{Vancouver}
  \country{Canada}
}

\author{Samuel Dodson}
\affiliation{%
  \institution{University at Buffalo}
  \city{Buffalo}
  \country{United States}
}

\author{Dongwook Yoon}
\affiliation{%
  \institution{The University of British Columbia}
  \city{Vancouver}
  \country{Canada}
}

\newcommand{\anchoredai}{\emph{AnchoredAI}}

\begin{abstract}
\emph{\textbf{ABSTRACT}} Generative AI is increasingly integrated into writing support, yet current chat-based interfaces often obscure referential context and risk amplifying automation bias and overreliance. We introduce \anchoredai, a novel system that anchors AI feedback directly to relevant text spans. \anchoredai{} implements two key mechanisms: (1) an Anchoring Context Window (ACW) that maintains unique, context-rich references, and (2) an update-aware context retrieval method that preserves the intent of prior comments after document edits. In a controlled user study, we compared \anchoredai{} to a chat-based LLM interface. Results show that \anchoredai{} led to more targeted revisions while fostering a stronger agency metrics (e.g., control and ownership) among writers. These findings highlight how interface design shapes AI-assisted writing, suggesting that anchoring can mitigate overreliance and enable more precise, user-driven revision practices.
\end{abstract}

\keywords{Collaboration, Education, Learning, LLM}

\maketitle

\fancyfoot{} 
\renewcommand\footnotetextcopyrightpermission[1]{}

\section{Introduction}

Anchored commenting---where feedback is directly attached to specific text spans---has been a foundational interaction paradigm in CHI and CSCW research, enabling contextualized collaboration across diverse application domains. From early systems like PREP editor \cite{neuwirth1990issues}, XLibris \cite{beyond1998schilit}, and spatial hypertext frameworks \cite{spatialhypertext1995marshall} to modern application in NB \cite{Zyto-2012}, Korero \cite{korero2017}, and web annotation protocols \cite{kahan2001annotea}, anchored comments preserve the spatial and semantic relationship between feedback and its referents. This paradigm has proven particularly effective in creating collaborative reading structures through tight comment-referent coupling \cite{Marshall-1998}. The success of anchored feedback also stems from its affordance to maintain contextual grounding, support asynchronous collaboration, and enable fine-grained discussions about specific content \cite{ohara-2002, wojahn1998interfaces}.

AI writing tools are now widely used across academic, professional, and personal contexts. Surveys show rapid adoption, and recent advances in large language models (LLMs) suggest strong potential for delivering high-quality, context-aware feedback \cite{sidoti-2025teens, sidoti-2025adults, hai-ai-index-report-2025}. However, the recent integration of generative AI into writing support has predominantly adopted \emph{chat-based interfaces} that present feedback in isolated conversational threads, departing form established commenting paradigms. In these interfaces, AI feedback is delivered as general suggestions or passage-level rewrites. While this format is familiar because it reflects how most AI chatbots are currently designed, it requires writers to map feedback to its intended location, increasing cognitive effort and reducing precision. For example, if a writer asks an AI to ``identify verb tense disagreements,'' the system may return a list of issues in a chat window without linking them to specific sentences. Writers must then search for each instance, risking missed errors or inconsistent fixes. While some popular AI-based writing tools, such as ChatGPT Canvas and Grammarly, offer anchored feedback, they either require users to manually select text for revision or are limited to detecting only basic errors. These tools also lack conversational support for discussing feedback in context and struggles to track changes to the text over time (elaborated further in Section~\ref{AI-Assisted-Writing}).

Moreover, the disconnection between AI feedback and document context risks amplifying automation bias and overreliance, potentially diminishing writers' agency over the writing process. Recent studies demonstrated that AI explanations can sometimes increase uncritical acceptance rather than promoting appropriate reliance \cite{bansal2021}. The lack of spatial grounding exacerbates these risks---when feedback is presented at the level of entire passages or full documents, writers lose contextual cues necessary for critical evaluation, may bypass close reading and critical engagement by copying and pasting AI-generated text. Over time, this dynamic can shift the AI's role from supportive collaborator to directive author, reducing the writer's agency and deep engagement with their own work \cite{kosmyna-2025}. These challenges point to an opportunity: as shown in \cite{Bucinca-2021}, structured interaction design (e.g., cognitive forcing functions) can reduce overreliance by nudging deeper engagement, and similarly the precision and clarity of anchored commenting may promises mitigating such agency risks in AI-assisted writing.

This paper explores how integrating anchored commenting into AI writing support can address these challenges and better align with the types of collaborative writing practices described in previous HCI studies \cite{birnholtz2012tracking, kim2001reviewing, neuwirth1990issues, wang2017write, weng2004asynchronous}. In this context, our study asks:

\begin{itemize}
  \item RQ1 (System/Design): How do we design an LLM pipeline for anchored AI comments? What technical challenges arise (e.g., referent ambiguity, robust re-anchoring) and how can they be addressed?

  \item RQ2 (Behavior): How does the format of AI feedback (chat vs. anchored) influence writers' editing behaviors and interaction patterns?
    
  \item RQ3 (Agency): How does presenting AI feedback as anchored comments (vs. chat) shape writers' agency, measured in perceived control, authorship, and ownership?
\end{itemize}

To answer RQ1, we introduce \anchoredai, a system that combines the precision of anchored comments with the generative capabilities of LLMs by attaching suggestions directly to their referent spans. \anchoredai{} allows writers to issue document-wide queries, such as ``Identify locations of verb tense disagreements,'' and receive feedback positioned exactly where it is needed. Two technical mechanisms underpin \anchoredai: (1) an Anchoring Context Window (ACW) maintains unique references by ensuring each AI comment preserves optimally sufficient contextual information to remain identifiable even as surrounding text changes, and (2) an update-aware context retrieval method preserves comment intent after document edits, leveraging semantic understanding to maintain the original purpose of AI suggestions across document revisions.

To address RQ2 and RQ3, we ran a controlled user study comparing \anchoredai{} to a chat-based LLM interface, examining how the presentation format of AI feedback influences both writer's revision behaviors and their subjective experiences of agency, finding that anchoring was associated with more targeted revisions, greater perceptions of agency, control, and ownership than Chat, although using the Anchored interface required more cognitive effort.

This study makes the following contributions to HCI, with a particular emphasis on AI-assisted writing and anchored feedback interfaces:

\begin{itemize}
    \item (System) \anchoredai, a novel system consisting of technical pipeline that anchors AI-generated feedback directly to text spans through an ACW and update-aware context retrieval, and
    
    \item (Empirical) The first controlled comparison of anchored vs. chat-based AI feedback interfaces, demonstrating that anchored presentation leas to more targeted revision, significantly enhanced agency markers (i.e., perceived control and ownership,) though with moderate increase in cognitive effort.
\end{itemize}

These establish a foundation for designing interfaces that promote critical engagement over passive acceptance and support precise, user-driven revision in AI-assisted writing.

The paper proceeds with a review of related work on anchored commenting, collaborative writing, and AI-assisted writing. It then introduces the technical design of our LLM-based anchoring methods, followed by a description of \anchoredai, a prototype system for generating anchored feedback. Next, we outline our evaluation methodology and present findings from a controlled user study comparing chat-based and anchored feedback interfaces. We conclude by discussing how anchored AI feedback supports writers’ agency, control, and ownership, and offer design implications for future AI writing tools.

\section{Related Work}

\subsection{Benefit of Anchored Commenting}

Anchored commenting builds on the long-standing practice of active reading. Active reading refers to the process of engaging with a text through annotation, questioning, and summarization to deepen understanding of the text. Annotations often include writing activities such as summarizing claims, highlighting evidence, and noting areas of uncertainty. Through this process, readers engage with the text in ways that are similar to the work of writers and editors during revision. In collaborative writing, similar actions occur: writers and reviewers locate problems, suggest changes, and explain their reasoning. As Marshall notes, through active reading practices, ``the roles of reader and writer blur'' (p.~131), as readers externalize their interpretations and begin to shape the meaning of the text \cite{Marshall-1998}.

Research on annotations have highlights their many forms, and how these enable them to refer to specific content in a text. Marshall's work on textbook annotation examined the marks that college students make in their textbooks \cite{Marshall-1997}. These annotations included underlines, highlights, marginal notes, and symbols. Marshall observed that readers often used visual cues such as arrows, brackets, and spatial proximity to connect their comments to specific text segments. In later work, she described these annotations as ``hypertextual'' because the visual cues acted as anchors, pointing from the comment to its referent in the text \cite{Marshall-1998}. Thus, \emph{anchored} commenting refers to feedback that is explicitly linked to a specific span of text, such as a word, phrase, or sentence, rather than provided in a separate location. Without these anchors, comments lack clear referents, making interpretation difficult.

Studies of collaborative writing further demonstrate that anchored feedback improves both understanding and revision. Anchoring reduces ambiguity by explicitly connecting a comment to its relevant text, which lowers cognitive effort for the writer \cite{ohara-2002}. Adler and colleagues found that professionals spend up to half their document time writing notes and annotations, underscoring the centrality of these practices in collaborative work \cite{adler-1998}. Studies of classroom-based learning have similarly found that anchored comments accelerate peer review and improve writing outcomes \cite{zhou-2012, Zyto-2012}. It is not surprise then that research and guidelines on the design of digital reading environment interfaces  emphasize the importance of anchored comments \cite{marshall2009reading, pearson2014designing, tashman2011active}.

When comments are anchored to specific text, writers no longer need to search for context or guess where feedback applies. Instead, they can focus their attention on making meaningful revisions. Anchored comments transform feedback from a vague suggestion into a precise, actionable cue, allowing writers to spend less time interpreting and more time improving their work. The specificity of a referent that comes with an anchored comment is especially important in writing, where revision is fundamentally an act of active reading. Writers must critically examine their own text, identify problems, and decide how to address them.  Reviewers do the same when providing feedback. Anchored comments extend the benefits of active reading into collaborative writing by making feedback explicit, localized, and easier to act upon. Without anchors, comments risk becoming vague or misinterpreted, which increases cognitive effort and reduces the quality of revisions.

\subsection{Collaborative Writing}

Collaborative writing involves two or more individuals producing a shared text through iterative cycles of drafting, reviewing, and revising. Prior research has shown that collaborative writing encompasses a range of activities, including brainstorming, researching, planning, composing, editing, and reviewing \cite{posner1992people, suchman1986framework}. These activities unfold across both synchronous and asynchronous modes. Writers may co-author content in real time, while commenting and proofreading often occur asynchronously \cite{kim2001reviewing}. Long-term projects frequently shift between modes, using comments for asynchronous editing, synchronous discussion for complex decisions, and comments again for final revisions \cite{boellstorff2013friends, olson2017people}. The increasing availability of online writing platforms such as Google~Docs and Microsoft~Word has made these workflows more common and accessible. These tools support real-time co-authoring, commenting, and version tracking.

Comments play a critical role in coordinating these efforts \cite{sellen1997paper}. By linking discussion directly to text, comments mark what needs attention and why, making feedback easier to act on during revision \cite{birnholtz2012tracking, churchill2000anchored, park2023why, weng2004asynchronous}. Comments also serve as signals of the work that has been done and what remains, supporting awareness among collaborators. Other awareness features, such as presence indicators, revision histories, and notifications support coordination by showing who is active, what has changed, and when edits occur \cite{dourish1992awareness, tam2006framework}. Tools like DocuViz visualize the revision history of a document, helping writers understand how contributions evolve over time \cite{wang2015docuviz}. These systems reinforce awareness by making the writing process more transparent and traceable.

Interface design also plays a critical role in shaping how collaborators communicate about writing. Wojahn et al. found that the structure of annotation interfaces affect both the number and types of tasks collaborators discuss during revision \cite{wojahn1998interfaces}. Neuwirth and colleagues extended this work by exploring how voice annotations support richer expression of feedback in distributed writing environments, showing that more expressive modalities can improve the usability of feedback \cite{neuwirth1994modalities}. Bietz found that the medium through which comments are delivered also affects how writers interpret its tone and credibility. Writers have been found to be less likely to accept critical feedback when it was delivered through text alone, compared to video or voice, highlighting the importance of perceived legitimacy and expressiveness in reviewer communication \cite{bietz2008feedback}.

In addition to the content of a comment, social dynamics shape how collaborators engage with the text and make sense of comments. Birnholtz and Ibara examined how tracking changes in collaborative documents affects not only the text but also the social dynamics between co-authors \cite{birnholtz2012tracking}. Co-authors who lack trust often prefer commenting over direct editing, while trusted collaborators are more likely to make edits themselves \cite{birnholtz2012tracking, birnholtz2013write}.

\subsection{AI-Assisted Writing}
\label{AI-Assisted-Writing}

HCI research has increasingly examined how LLMs can support writing across a wide range of contexts, including creative writing, academic writing, everyday communication, and personal journaling \cite[e.g.,][]{chung2022talebrush, gero2022sparks, li2024value, kim2024diarymate, pang2025llm}. Early systems such as Smart Compose in Gmail offered phrase-level suggestions to reduce repetitive typing, demonstrating how AI could streamline routine writing tasks \cite{chen2019gmail}. As more powerful models like ChatGPT became widely available, researchers began examining their role in additional writing workflows, including ideation, revision, and narrative development.

In creative writing, AI has been explored as a collaborator that can generate stylistic rewrites, plot structures, and dialogue. Tools like Wordcraft and Dramatron allow writers to issue open-ended prompts and receive varied textual outputs, supporting exploration and experimentation \cite{yuan2022wordcraft, mirowski2023dramatron}. These systems position AI as a co-author or ideation partner, helping writers overcome creative blocks and consider new directions. Singh et al. found that multimodal suggestions could facilitate integrative leaps, enabling writers to connect disparate ideas more effectively. Similarly, Dang et al. developed systems that generate paragraph-level summaries and annotations, offering writers an external perspective on their drafts and supporting more reflective revision practices \cite{dang2022beyond}.

Despite these benefits, the integration of AI into writing raises concerns about authorship, control, and psychological ownership. Biermann et al. identified three key barriers to adoption: writers' emotional investment in the act of transforming ideas into words, skepticism about AI's ability to handle nuanced tasks such as character development, and mismatches between system control mechanisms and individual writing strategies \cite{biermann2022companion}. These concerns reflect a broader tension in design: while AI can make writing easier, it may also reduce the writer's sense of agency and engagement. Importantly, these groups should not be treated as monolithic. As Kobiella and colleagues' diary study of young professionals shows, individuals within the same demographic can have divergent experiences with AI-based writing tools. Some participants reported increased creative output and satisfaction, while others described diminished ownership and perceived mediocrity \cite{kobiella2024if}.

Recent studies have examined how prompt design influences psychological ownership. Joshi and Vogel investigated whether writing with generative AI diminishes this sense of ownership and found that longer prompts were associated with higher levels of personal investment \cite{joshi2025ownership}. Participants who wrote more detailed prompts reported thinking more deeply about their stories, although the benefits plateaued when prompt length approached the full story length.

The influence of AI on content itself is another area of concern. Jakesch et al. demonstrated that ``opinionated'' language models—those that favor certain viewpoints—can subtly shape user output, raising questions about neutrality and authorship \cite{jakesch2023opinionated}. As AI-generated content becomes more prevalent, users may experience a diminished sense of control over their work. Control refers to the ability to direct the content, tone, and structure of a piece \cite{biermann2022companion}, while ownership reflects emotional investment and responsibility for the final product. Although experienced professionals often use AI to enhance productivity, younger professionals have reported reduced feelings of accomplishment, citing insufficient challenge and a perception of lower-quality outcomes \cite{kobiella2024if}.

Li et al. documented both the benefits and risks of generative AI in writing. While users reported increased productivity and confidence, they also expressed concerns about reduced accountability and diminished diversity in writing styles \cite{li2024value}. These findings underscore the central design tension: AI can support writing, but it may also erode the writer's sense of agency and authorship. We argue that such tensions are necessary for understanding the design spaces for AI-based writing assistants \cite{lee2024design}. This tension also reflects how AI workflows may evolve in response to the needs, expectations, and values of writers. Guo et al. found that creative writers' engagement with AI is shaped by values of authenticity, ownership, creativity, and craftsmanship \cite{guo2025pen}. Kim et al. further raise questions about authorial control by exploring how AI-generated texts can be adapted not only by writers but also by audiences \cite{kim2024authors}.

Questions of attribution and credit also emerge when AI contributes to the writing process. He et al. found that while users recognize AI's contributions in co-creative writing, they consistently assign it less credit than human collaborators, with attribution shaped by contribution type, initiative, and personal values \cite{he2025credit}. This tension is further illustrated by the ``AI ghostwriter'' effect. Draxler et al. found that as AI contributions increased, participants reported lower feelings of ownership, even though they continued to identify themselves as the authors \cite{Draxler-2024}. A follow-up study revealed that participants were more willing to credit human ghostwriters than AI, suggesting that reliance on AI reduces perceived ownership without necessarily changing public authorship claims.

However, interaction design can help mitigate these effects. Wasi and colleagues found that when users actively edited or selectively incorporated AI suggestions, their feelings of control and accountability improved \cite{Wasi-2024}. These findings suggest that thoughtful design can support meaningful engagement, allowing writers to maintain a sense of authorship even in highly AI-supported workflows.

Some AI-based writing tools, such as ChatGPT's Canvas, offer anchored feedback by allowing users to select specific spans of text and request suggestions. However, this approach requires the writer to first identify which parts of the document need revision. As a result, the tool shifts a key aspect of the reviewing process---i.e., detecting areas for improvement---back onto the writer. Canvas therefore provides limited proactive support, as it cannot independently identify and anchor feedback to relevant portions of the text. Grammarly, another widely used AI writing assistant, can automatically detect errors and offer anchored suggestions. Yet its feedback is mostly limited to surface-level issues such as grammar and spelling, which restricts its usefulness to copyediting. Neither Canvas nor Grammarly supports conversational feedback, such as explaining the rationale behind a suggestion, despite research indicating that such explanations help writers decide whether to accept a revision \cite[e.g.,][]{park2023why}. Additionally, both tools lack functionality for tracking changes over time, which reduces their effectiveness in supporting complex and iterative writing tasks in which the text may be written and rewritten many times.

Previous research has shown that anchored commenting supports collaborative revision by linking feedback directly to specific parts of a text. At the same time, most AI-assisted writing tools present feedback in chat-based formats that lack clear connections to the text. Furthermore, while large language models can support ideation and revision, they may also reduce writers’ sense of agency, ownership, and control. Studies have emphasized the importance of interaction design in shaping how writers engage with AI, suggesting that more grounded and context-aware feedback can help preserve authorship and promote deeper engagement. Building on these insights, this research investigates how integrating anchored commenting into AI writing tools can address these concerns and better support collaborative writing practices that prioritize clarity, control, and user agency.

\section{LLM Anchoring Techniques for \anchoredai}

The core technical contribution of \anchoredai{} is a pipeline designed to robustly generate and maintain anchored AI feedback across two primary interaction modes. The first mode involves user-initiated \emph{meta-commentary}, where the system responds to a user's document level query by generating and anchoring multiple comments throughout the text. The second mode supports user-initiated \emph{text-anchored comments}, where they select a specific text span to engage in a contextual dialogue with the AI. Both modes must overcome the fundamental technical challenges of anchor persistence, to ensure that comments remain correctly attached and contextually relevant even when the underlying document is revised. We detail the mechanisms that address these challenges, including our structured anchor generation process, the Anchoring Context Window (ACW) for resolving referent ambiguity, and an update-aware retrieval method for maintaining conversational context.

\subsection{AI Response of User's Meta-commentary}
\label{sec:overall_pipeline}

The first interaction mode, which we termed \emph{user-initiated meta-commentary} (or meta-commentary in short), allows users to issue document-wide queries that require AI feedback at multiple locations. For example, a user might ask the system to ``identify locations of verb-tense disagreements'' or ``suggest more concise phrasing for redundant sentences.'' This requires the LLM to not only identify relevant text spans but also return them in a structured output data that the editor can parse reliably, bridging the gap between a LLM's conversational output and required formality of the parsing engine.

\subsubsection{Overall Pipeline}
\label{sec:pipeline1}

\begin{figure}
    \centering
    \includegraphics[trim=0 250 0 100,clip,width=1.0\columnwidth]{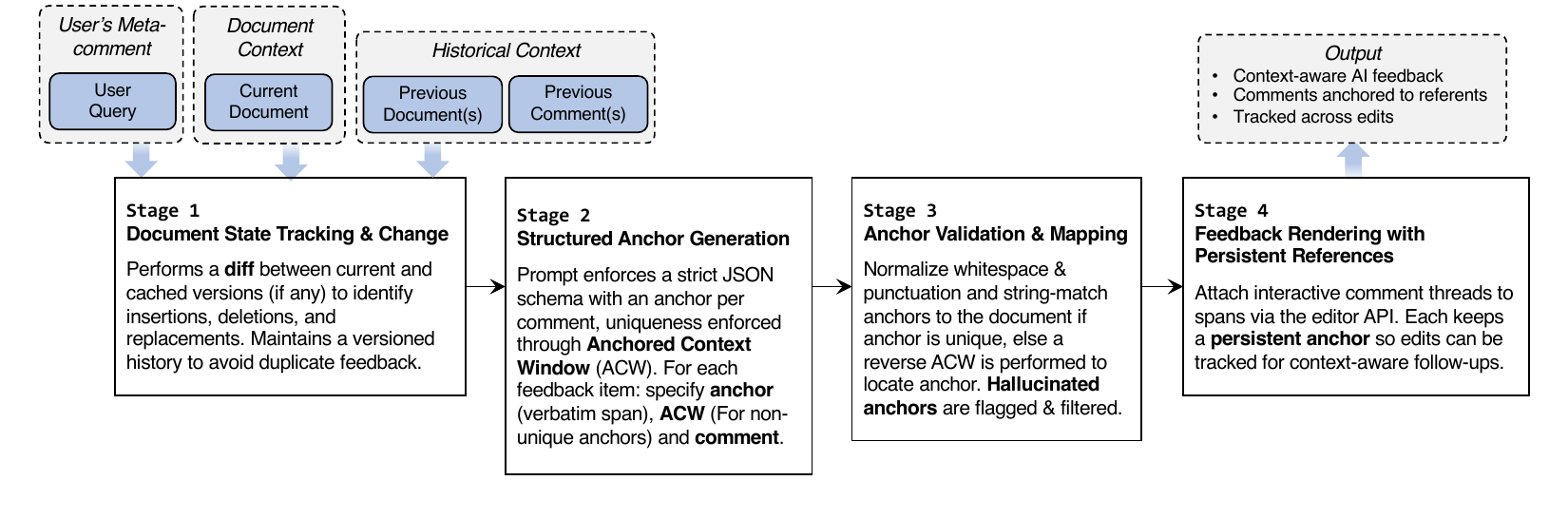}
    \caption{Procedure of the anchoring mechanism.}
    \Description{Diagram of the anchoring mechanism for AI-assisted feedback in a four-stage process. The system begins by tracking document state changes, comparing current and any cached versions to detect insertions, deletions, and replacements while maintaining a version history. Next, feedback is generated in a JSON format with one anchor per comment, ensuring uniqueness through an Anchored Context Window (ACW). Then, anchors are validated and mapped by normalizing whitespace and punctuation characters, string-matching to the document, and performing reverse ACW searches if necessary. Hallucinated anchors are filtered out. Finally, feedback is rendered as interactive comment threads that remain persistently anchored to document spans, enabling comments to stay linked across edits and allowing for context-aware follow-ups. The result is AI feedback that is both context-aware and reliably tracked throughout document revisions.}
    \label{fig:metacommenting}
\end{figure}

The \anchoredai{} workflow for meta-commentary consists of four key stages that transform user queries into precisely-located, contextually-aware feedback:

\begin{enumerate}
    \item \emph{Document State Tracking and Change Detection.} When a user submits a query, the system first performs a differential analysis between the current document state and the cached previous version. This comparison identifies all modifications---insertions, deletions, and replacements---ensuring the AI understands what has changed since the last interaction. The system maintains a versioned document history including previous comments to prevent feedback duplication and enable context-aware responses.

    \item \emph{Structured Anchor Generation.} The user's query, along with the updated document and change history, is processed through a carefully engineered prompt that enforces a strict JSON schema and a unique anchor. This schema mandates that the LLM must specify, for each piece of feedback it generates: (a) the exact text span from the document to anchor on, (b) the ACW (See Section \ref{ACW_description}) if the selected text span is not unique, and (c) the comment content. The prompt explicitly instructs the model to extract verbatim text spans from the current document when determining anchor points, reducing the likelihood of hallucinated anchors.

    \item \emph{Anchor Validation and Mapping.} Each LLM-generated anchor undergoes validation against the actual document content using string matching with normalization for whitespace and punctuation variations. The system attempts to locate each anchor text within the document. For none unique anchors, ACW returned from the LLM will be used to determine the location of referred referent location. Hallucinated anchors that cannot be mapped to existing text are flagged as invalid and filtered out, ensuring only grounded feedback reaches the user.

    \item \emph{Feedback Rendering with Persistent References.} Valid anchors are transformed into interactive comment threads attached to their corresponding text spans using the document editor's API. Each comment maintains a persistent reference to its anchor location, enabling the system to track whether the anchored text has been modified or deleted in subsequent edits. This persistence allows for context-aware follow-up interactions within comment threads (Section~\ref{sec:context-awareness}).
\end{enumerate}

\subsubsection{Resolving Ambiguous References with Optimally Unique Anchoring Context Windows (ACWs)}
\label{ACW_description}

\begin{figure}
    \centering
    \includegraphics[trim=0 250 0 100,clip,width=1.0\columnwidth]{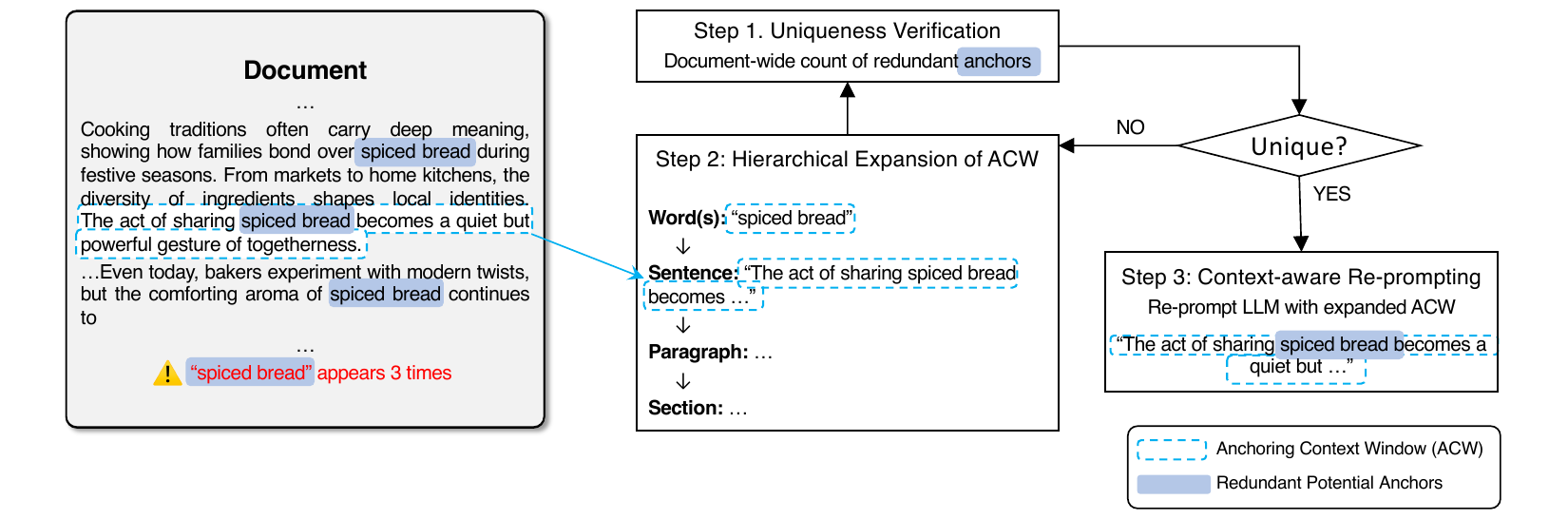}
    \caption{Anchoring Context Window (ACW) adjustment. When a non-unique anchor is selected, the system dynamically expands from word $\rightarrow$ sentence $\rightarrow$ paragraph until the context is unique.}
    \Description{Diagram of how the system resolves ambiguous or redundant anchors in a document. First, the system performs verifies the uniqueness of the anchor by checking whether an anchor appears multiple times. If the anchor is not unique, the system engages in hierarchical expansion of the ACW, progressively enlarging the span from the word to the containing sentence, paragraph, or section until a unique reference is identified. Once uniqueness is achieved, the system performs context-aware re-prompting, sending the expanded ACW back to the language model to ground the feedback in the precise location of the document. This process ensures that feedback remains tied to a uniquely identifiable passage, avoiding confusion caused by repeated terms.}
    \label{fig:ACW_Process}
\end{figure}

\begin{figure}
    \centering
    \includegraphics[trim=0 110 0 100,clip,width=1.0\columnwidth]{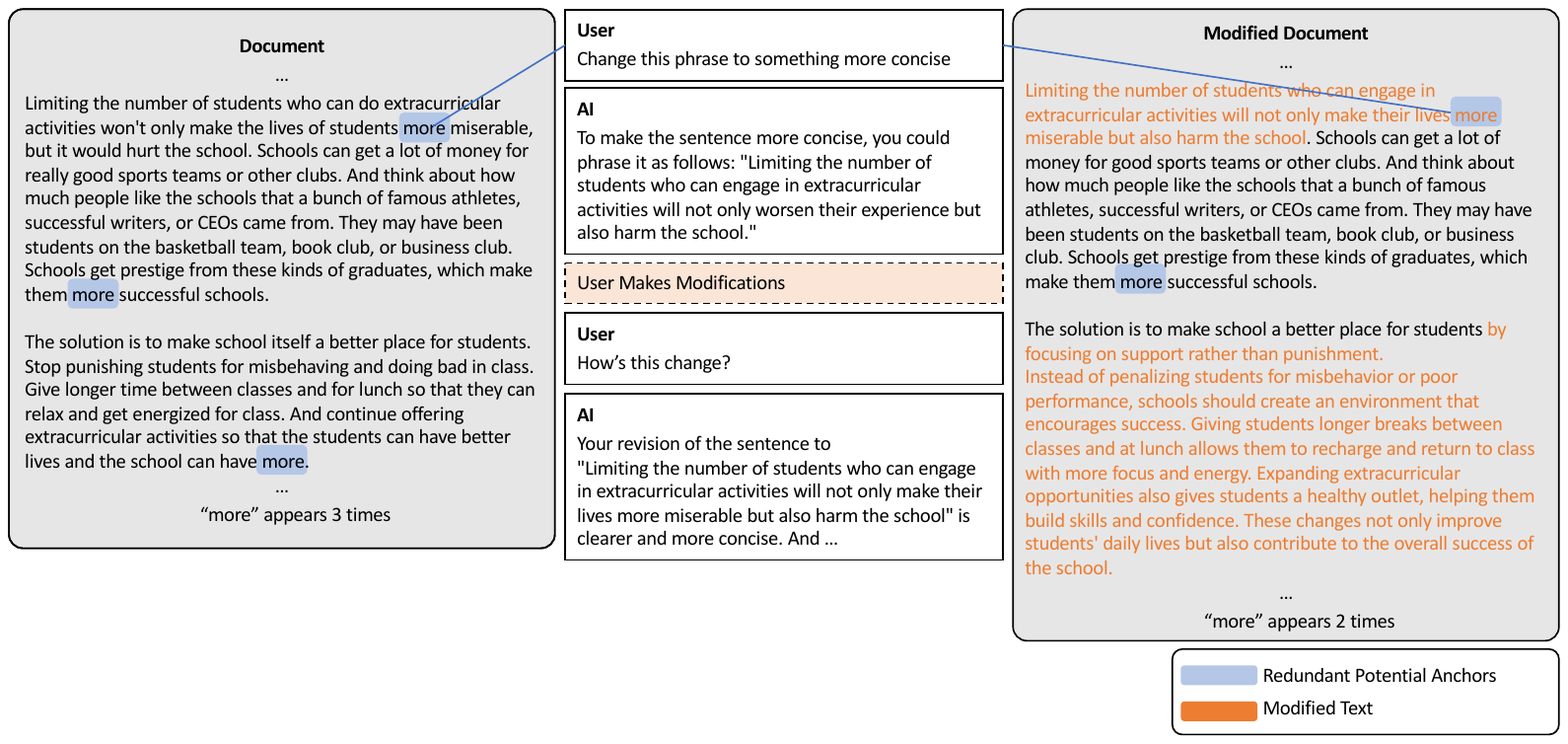}
    \caption{Example of a user-initiated anchored comment thread. By applying the Anchoring Context Window (ACW), the LLM resolves ambiguity when the referent is a non-unique word and after subsequent document modifications, understand the modification referred to by the user.}
    \Description{Diagram of how AI-generated feedback is tied to specific text spans and persists after the user revises the document. On the left, the original document includes multiple redundant instances of the word `more.' The user asks for a more concise phrasing, and the AI suggests a rewritten sentence. After the user accepts and edits the text, the modified document (right) highlights the updated span in orange while redundant anchors remain tracked in blue. The interaction flow illustrates how the anchoring mechanism supports precise comment alignment, enabling feedback to remain valid and trackable even after text changes.}
    \label{fig:ACW_Example}
\end{figure}

A key issue with anchored comments is the possibility of a non-unique referent. When an anchor is selected, it may select on minimal text spans---such as single words or short phrases---as anchor points. These selections often appear multiple times throughout a document, creating ambiguity about where the comment should be attached. Simply rejecting such anchors would result in lost feedback, while arbitrarily choosing the first occurrence could misplace important suggestions.

Our solution introduces \emph{Anchoring Context Window (ACW) adjustment}, a recursive expansion algorithm that transforms ambiguous anchors into unique, precisely-locatable references (see Figure~\ref{fig:ACW_Process}). The algorithm operates as follows:

\begin{enumerate}
    \item \emph{Initial Anchor Analysis:} When an comment is established on an anchor, the system first checks if this text appears uniquely in the document. If unique, it proceeds directly with this anchor.

    \item \emph{Hierarchical Context Expansion:} If the selected anchor is non-unique, the system expands to progressively larger semantic units: (1) \emph{Word $\rightarrow$ Sentence}: Expand to the complete sentence containing the selection, (2) \emph{Sentence $\rightarrow$ Paragraph}: Expand to the full paragraph if the sentence remains ambiguous, and (3) \emph{Paragraph $\rightarrow$ Section}: Continue expanding to larger structural units as needed.

    \item \emph{Context-aware Re-prompting with Expanded ACW:} If expansion was necessary, the system re-prompts the LLM with the expanded ACW, allowing it to refine its feedback with awareness of the fuller context. This ensures the comment remains relevant to the intended location despite the expanded anchor.
\end{enumerate}

This mechanism ensures that even when a minimal anchor point is selected, the resulting feedback is precisely located and contextually appropriate, as illustrated in Figure~\ref{fig:ACW_Example}.

\subsection{AI Responses to User's Text-anchored Comments}
\label{sec:context-awareness}

Beyond generating responses to users' meta-commentary, \anchoredai{} supports a more targeted user interaction where they can select a specific text span and engage in a contextual dialogue with the AI. This mirrors the familiar workflow of leaving a comment for a human collaborator, when discussing on specific constructs, sentences, or phrases. The central challenge in this mode is to maintain conversational context and referential integrity when the anchored text is modified by the user between human-AI conversation turns. For example, after receiving an AI's suggestion to rephrase a sentence, a user might implement the change and then ask, ``Is this version better?'' Then, the system must recognize the change, understand the nature of the edit, and relate it to the prior conversation. Our approach addresses this by localizing document changes relative to the anchor and providing the LLM with this dynamic context.

\subsubsection{Overall Pipeline}

To enable AI's response to user's text-anchored comments, we extended the pipeline for meta-commentary in Section~\ref{sec:pipeline1}. This pipeline supports richer interaction by enabling users to chat within anchored comment threads or initiate new ones. Each anchored comment thread is treated as its own chat interface: for every thread, the system provides the model with the full document, the Anchoring Context Window (ACW) if the referent is non-unique, and the initial comment. This design ensures that any follow-up discussion remains grounded in the relevant context of the comment and the surrounding text.

\begin{figure}
    \centering
    \includegraphics[trim=0 250 0 100,clip,width=1.0\columnwidth]{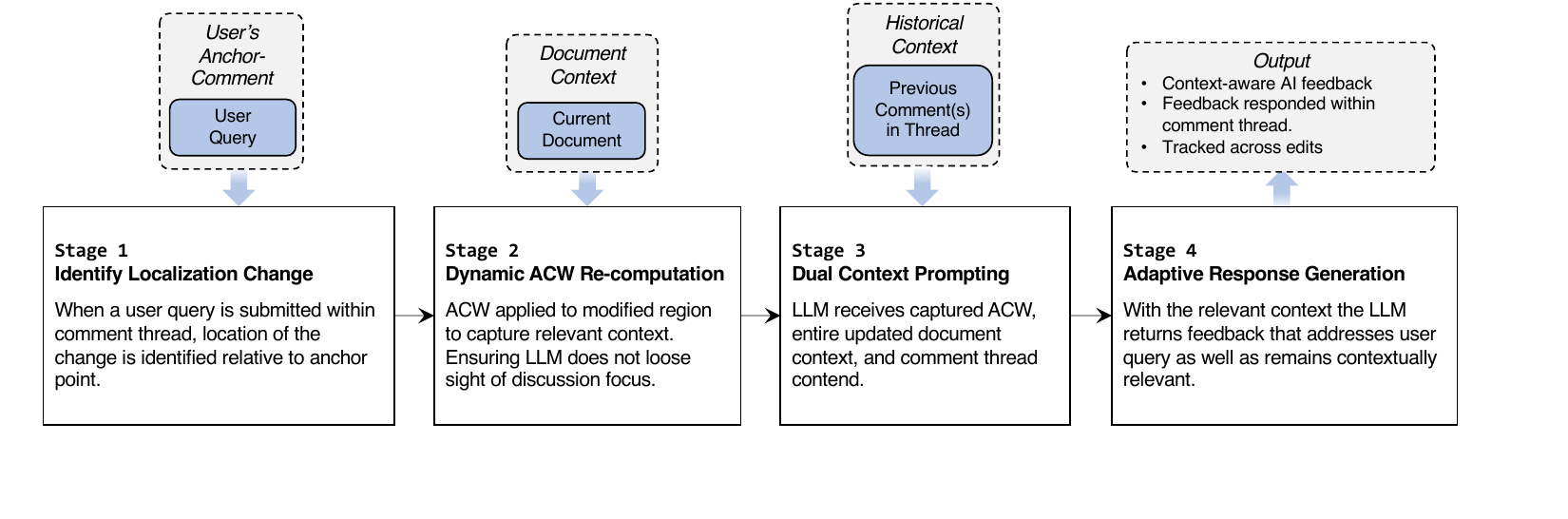}
    \caption{Procedure of the anchored commenting thread mechanism.}
    \Description{Diagram of the four stages for processing user queries within anchored comment threads. First, localization change is identified by mapping the user's query to the anchor point in the document. Second, dynamic Anchoring Context Window (ACW) re-computation is applied to the modified region to ensure relevant context is captured without losing discussion focus. Third, the system performs dual context prompting, in which the language model receives the updated ACW, full document context, and prior comment thread content. Finally, adaptive response generation occurs, producing feedback that addresses the user's query while remaining contextually relevant and persistently anchored, ensuring feedback is tracked across edits.}
    \label{fig:anchorcommenting}
\end{figure}

\subsubsection{Challenge and Solution: Resolving Ambiguous References with Optimally Unique Anchoring Context Windows (ACWs).}

When users modify their content and then return to previous comment threads for follow-up discussion, the AI must understand both what has changed and how those changes relate to its earlier feedback. Simply resubmitting the entire updated document would overwhelm the model and dilute focus from the specific modification relevant to the anchored discussion.

Our approach combines \emph{targeted context provision with full document awareness}:

\begin{enumerate}
    \item \emph{Change Localization:} When a user engages with an existing comment thread after editing, the system identifies the precise location and nature of changes relative to the original anchor point.
        
    \item \emph{Dynamic ACW Recomputation:} The system applies ACW adjustment (as described above) to the modified region, capturing enough context to encompass both the original anchor and the relevant changes. This ensures the model sees the complete local modification without losing sight of the specific discussion point.
        
    \item \emph{Dual-Context Prompting:} The LLM receives: (1) The targeted ACW containing the modification and surrounding context, (2) The complete updated document for global coherence checking, and (3) The conversation history within that specific comment thread.
    
    \item \emph{Adaptive Response Generation:} With this dual awareness, the model can intelligently determine whether to affirm that previous feedback remains valid despite changes, retract suggestions that are no longer applicable, provide updated guidance reflecting the new content, and acknowledge successful implementation of prior suggestions.
\end{enumerate}

This mechanism preserves conversational continuity while adapting to content evolution, enabling natural back-and-forth refinement within each anchored context. As shown by the response post modification in Figure~\ref{fig:ACW_Example}.

\subsection{Implementation Details}

The anchoring system is implemented as a client-server architecture with real-time document synchronization. The client interface leverages the Microsoft Word API for comment management, anchor tracking, and document state monitoring. AI responses are generated through OpenAI's GPT-4o API with custom prompt engineering for structured output generation. Document versioning, differential analysis, anchor validation and context window adjustment are performed client-side to minimize latency. The system maintains a persistent session store of document versions, comment histories, and anchor mappings to support seamless context preservation across editing sessions.

\section{Evaluation of \anchoredai}

\subsection{Hypothesis}

Previous research on collaborative writing suggests that the way feedback is presented can strongly shape how writers engage with revisions. In chat-based systems, feedback is often delivered in a conversational flow detached from specific passages of text. This can encourage writers to accept larger blocks of suggestions or to rework broader portions of their drafts, sometimes extending beyond what is strictly necessary. By contrast, anchoring feedback directly to text has the potential to ground revisions in localized context, encouraging users to focus on specific issues rather than rewriting broadly. From this reasoning, we hypothesize that anchoring will reduce large-scale rewriting in favor of smaller, more targeted revisions (H1).

Beyond influencing revision scope, the design of feedback interfaces can also affect writers' perceptions of agency and authorship. When suggestions are tightly coupled to the writer's own text, users may feel more in control of the revision process, more responsible for implementing changes, and more entitled to claim ownership over the final product. In contrast, feedback delivered through a chat interface may blur boundaries between the writer's contributions and the system's, potentially diminishing perceived authorship. Accordingly, we hypothesize that anchoring feedback to text will increase users' ownership, perceived sense of control, and authorship over their revised text (H2). Together, these hypotheses reflect our broader interest in how interface design mediates not only the mechanics of revision but also the writer's relationship to their work.

\subsection{Participants}

We recruited 22 university students between the ages of 18 and 24 through word of mouth and e-mail announcements. The sample included 10 self-identified males, 11 females, and one non-binary participant, representing 12 different academic majors. Most participants reported high English proficiency (20 identified as fluent or native speakers) and were comfortable using digital interfaces (21 indicated being very comfortable or comfortable). Participants received an honorarium for completing the study.

\subsection{Methods}

We conducted a within-subjects laboratory study to compare two AI feedback interfaces: a chat-based interface and an anchored-comment interface. Our goal was to understand how interface design influences both the effectiveness of AI-generated feedback and participants' perceptions of their role in the revision process.

To do this, participants revised multiple essays using both interfaces. Each revision task focused on either local edits (e.g., grammar, word choice, and style) or global edits (e.g., organization and argument structure). We assessed effectiveness through several measures: document snapshots and interaction logs captured participants' edits and behaviors, while copy-paste tracking provided insight into how feedback was incorporated. After each revision, participants completed surveys measuring perceived control, ownership, and cognitive workload. Finally, we conducted a semi-structured interview to gather qualitative feedback and better understand participants' strategies and experiences.

To isolate the effect of interface design, we restricted certain features of the anchored interface, such as user-initiated comments and within-thread conversations. This prevented participants from replicating a chat-like experience in the anchored condition, ensuring a clearer comparison between the two feedback formats.

\subsubsection{Experimental System}

\begin{figure}
    \centering
    \fbox{\includegraphics[trim=5 50 5 5,clip,width=1.0\columnwidth]{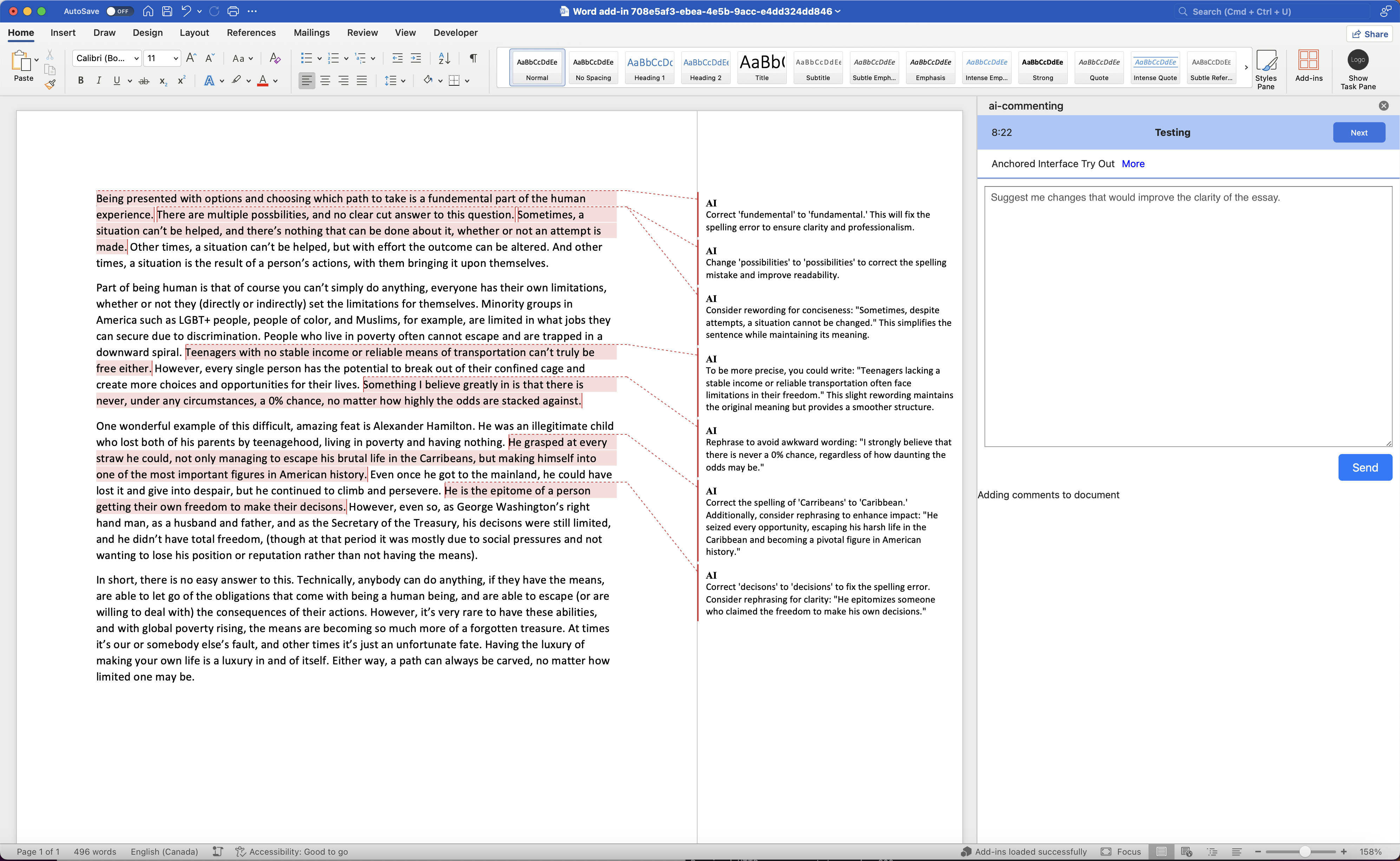}}
    \caption{Interface of \anchoredai{} Interface with comments.}
    \label{fig:anchor_ai_interface}
    \Description{Screenshot showing a Microsoft Word document with AI-generated comments anchored directly to highlighted text spans in the text. On the left, the document includes underlined portions where the AI suggestions are anchored. Red connector lines link each highlighted passage to corresponding AI feedback displayed in the right margin. The comments provide specific suggestions, such as spelling corrections, rephrasing for clarity, and style adjustments. On the far right, a sidebar titled ``ai-commenting'' includes a text box where the user can enter queries (e.g., ``Suggest me changes that would improve the clarity of the essay'').}
\end{figure}

Participants revised texts in Microsoft Word using a custom-built add-in that displayed AI feedback in one of two formats: anchored comments or chat messages. Aside from the feedback format, all other interface elements were identical across conditions.

The add-in also logged participants' interactions. Specifically, it recorded every copy-and-paste action, including the clipboard content and its source. This, for example, allowed us to distinguish between text copied from the document versus copied from the feedback interface. In addition, the system captured a document snapshot every 10 seconds, enabling us to track revisions beyond copy-paste actions.

\subsubsection{Materials}

Each session was conducted on a laptop configured with Microsoft Word, the study add-in, and supporting software. Figure~\ref{fig:anchor_ai_interface} shows the anchored interface as presented to participants.

Participants revised four essays drawn from publicly available ACT writing samples \cite{essay-a, essay-b, essay-c, essay-d}. The ACT is a standardized college admissions test in the United States that includes a writing section requiring students to compose an argumentative essay under timed conditions. Each essay was pre-scored at level three on the ACT rubric \cite{act-rubric}, representing an average performance characterized by basic organization and language use, along with noticeable issues in grammar, word choice, and style. The essays were comparable in length, error density, and error types. We selected these texts because they reflect authentic academic writing tasks while providing a consistent baseline for revision. Using standardized essays ensured experimental control and minimized variability that would otherwise arise from differences in participants' own writing proficiency, topic familiarity, or disciplinary conventions.

To measure participants' perceptions of authorship and control, we have selectively adapted relevant survey items from Draxler et al. \cite{Draxler-2024}. Table \ref{tab:survey-questions} lists the items and their targeted constructs. We also assessed cognitive and physical workload using the NASA-TLX instrument \cite{nasa-tlx}, a validated and widely used measure in HCI research.

\begin{table}
    \small
    \centering
    \begin{tabular}{ ll } 
      \toprule
      \textbf{Question} & \textbf{Targeted Perception} \\
      \midrule
      Q1. I am the main contributor to the content of this revision. & Ownership\\
      Q2. I have made substantial contributions to the revision of the text. & Ownership \\ 
      Q3. I am responsible for at least part of the text. & Ownership\\
      Q4. It felt like I was in control of the revision during the task. & Sense of Control \\
      Q5. My name should appear as the reviewer of this essay. & Authorship\\ 
      \bottomrule
    \end{tabular}
    \caption{Questions asked in study survey.}
    \label{tab:survey-questions}
\end{table}

\subsubsection{Procedure \& Tasks}

Figure~\ref{fig:study_procedure} summarizes the study procedure. Sessions were conducted individually in a quiet room, with an experimenter present to provide technical assistance if needed. Each session lasted approximately one hour and followed four main steps:

\begin{figure}
    \centering
    \includegraphics[trim=0 270 0 150,clip,width=1.0\columnwidth]{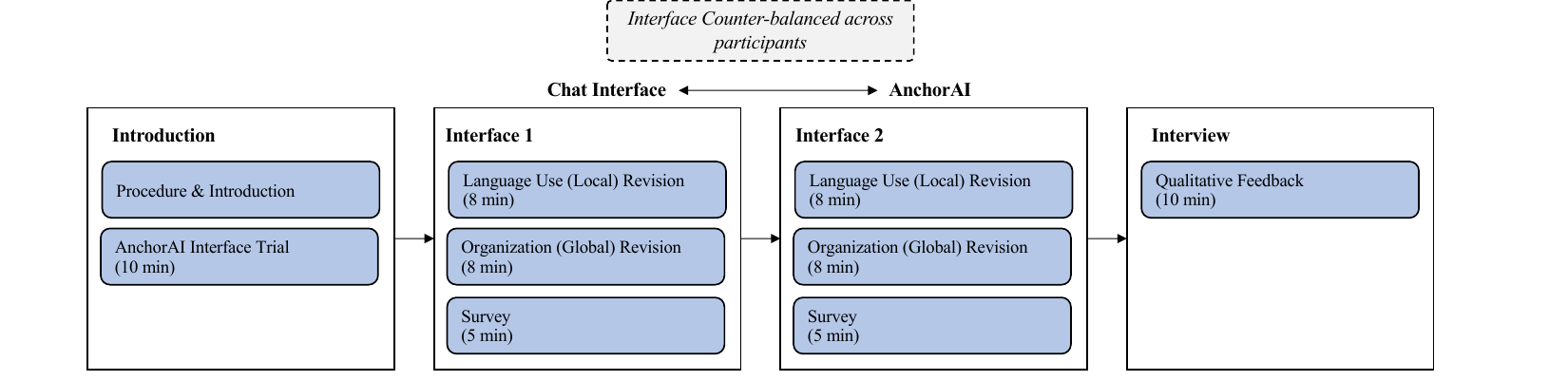}
    \caption{Procedure of \anchoredai{} Study.}
    \label{fig:study_procedure}
    \Description{Diagram of the study workflow, with tasks counterbalanced across participants between the Chat interface and the Anchored interface. The session begins with an introduction that includes procedure explanation and a 10-minute AnchoredAI trial. Participants then complete two interface conditions in sequence. In Interface 1, they conduct a local language-use revision (8 minutes), a global organization revision (8 minutes), and a short survey (5 minutes). In Interface 2, they repeat the same set of tasks (i.e., local revision, global revision, and survey) using the alternate interface. Finally, the session concludes with an interview for qualitative feedback.}
\end{figure}

\begin{enumerate}
    \item \emph{Introduction \& Familiarization:} Participants completed a consent form and demographic questionnaire. The experimenter then introduced the study and allowed participants up to 10 minutes to explore the \anchoredai{} interface. This familiarization period was optional; the study began once participants felt comfortable.
    
    \item \emph{Interface 1:} Participants were randomly assigned to start with either the chat interface or the anchored interface. In this block, they completed two timed revision tasks and a survey:
    
    \begin{itemize}
        \item \emph{Local Revision (8 minutes):} Participants revised an essay focusing on grammar, word choice, and style.
        
        \item \emph{Global Revision (8 minutes):} Participants revised a different essay focusing on argument structure, paragraph flow, and overall coherence.
        
        \item \emph{Survey (5 minutes):} After completing both tasks, participants filled out the perception survey and NASA-TLX.
    \end{itemize}
    
    The time limits in the local and global revision tasks ensured that all tasks were completed under comparable conditions, preventing differences in outcome quality from being attributed to unequal time spent. The eight-minute duration was chosen following pilot study feedback to strike a balance: it created enough time pressure to encourage efficiency while also allowing participants to read the essay, engage with the interface, and make substantive modifications to the text without experiencing fatigue or loss of focus.
    
    \item \emph{Interface 2:} Participants then switched to the other interface and repeated the same sequence of tasks with two new essays. Essay assignment was counterbalanced to control for content effects.
    
    \item \emph{Interview:} Finally, participants completed a semi-structured interview. The flexible format allowed experimenters to ask follow-up questions and clarify participants' reasoning, while also eliciting broader feedback on the UI/UX of each system and suggestions for improvement. We asked participants about their overall impressions of the experiment, whether certain types of mistakes in the text were overlooked by the LLM, and which interface they preferred and why. The interview also explored how participants' prompting strategies differed across the two interfaces. Importantly, this stage provided insight into why participants responded to survey items in particular ways, offering a deeper understanding of their revision strategies and perceptions of the two interfaces.
\end{enumerate}

\subsection{Findings}

Overall, participants completed the tasks successfully without significant usability or technical challenges. In the Chat condition, participants spent an average of 453.0 seconds ($SD = 85.8$) on Local revisions and 405.0 seconds ($SD = 119.0$) on Global revisions. They submitted an average of 2.7 queries ($SD = 1.8$), receiving lengthy AI responses that averaged 302.1 words ($SD = 107.6$). In the Anchored condition, task times were comparable at 456.4 seconds ($SD = 61.4$) for Local and 435.0 seconds ($SD = 93.9$) for Global revisions. This involved fewer queries (M = 1.9, $SD = 1.55$) and generated more granular feedback: an average of 14.5 concise comments ($SD = 7.9$) per task, with each comment average 28.6 words ($SD = 6.0$).

\begin{figure}
    \centering
    \includegraphics[trim=0 0 0 0,clip,width=1.0\columnwidth]{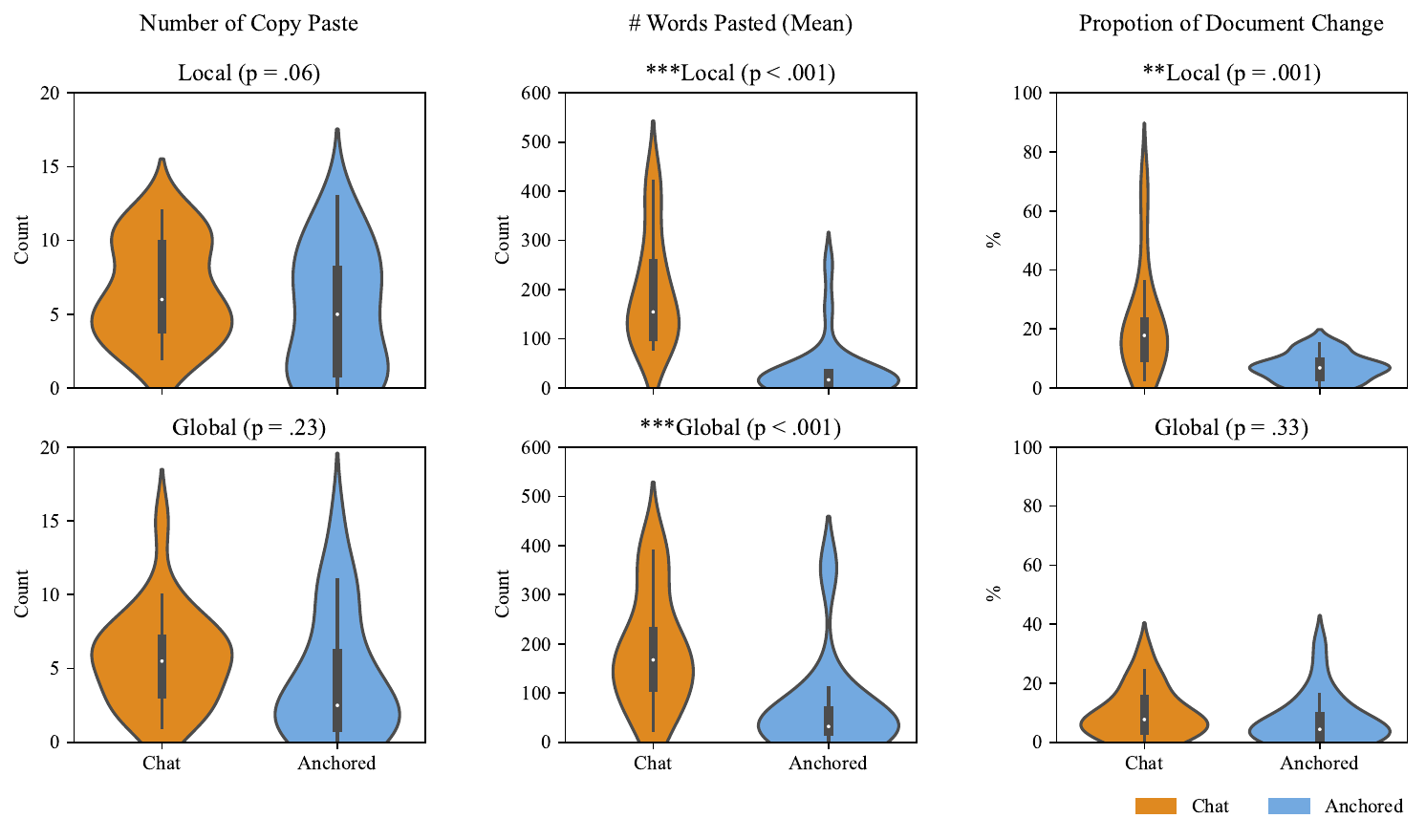}
    \caption{Distribution of revision behaviors across interface conditions. Violin plots show the density and variability of (left) number of copy-paste actions, (middle) mean number of words pasted per action, and (right) percentage of document changed. Each violin represents one condition: Chat \& Local, Chat \& Global, Anchored \& Local, and Anchored \& Global. Black bars indicate the interquartile range with the white dot marking the median.}
    \Description{Figure of violin plots comparing conditions across three measures: number of copy-paste actions, mean number of words pasted, and proportion of document changed. Results are shown separately for local and global revision tasks. For copy-paste actions, Anchored yielded fewer instances than Chat, with a trend toward significance in the local task (p = .06) but no difference in global tasks (p = .23). For words pasted, Anchored consistently resulted in significantly fewer words pasted than Chat for both local (p < .001) and global (p < .001) revisions. For proportion of document change, Anchored revisions produced smaller changes during local tasks (p = .001), while no difference was observed in global tasks (p = .33). Overall, the plots indicate that the Anchored interface encouraged more targeted revisions with less wholesale text replacement compared to the Chat interface.}
    \label{fig:revision-plot}
\end{figure}

\begin{figure}
    \centering
    \includegraphics[trim=0 0 0 0,clip,width=1.0\columnwidth]{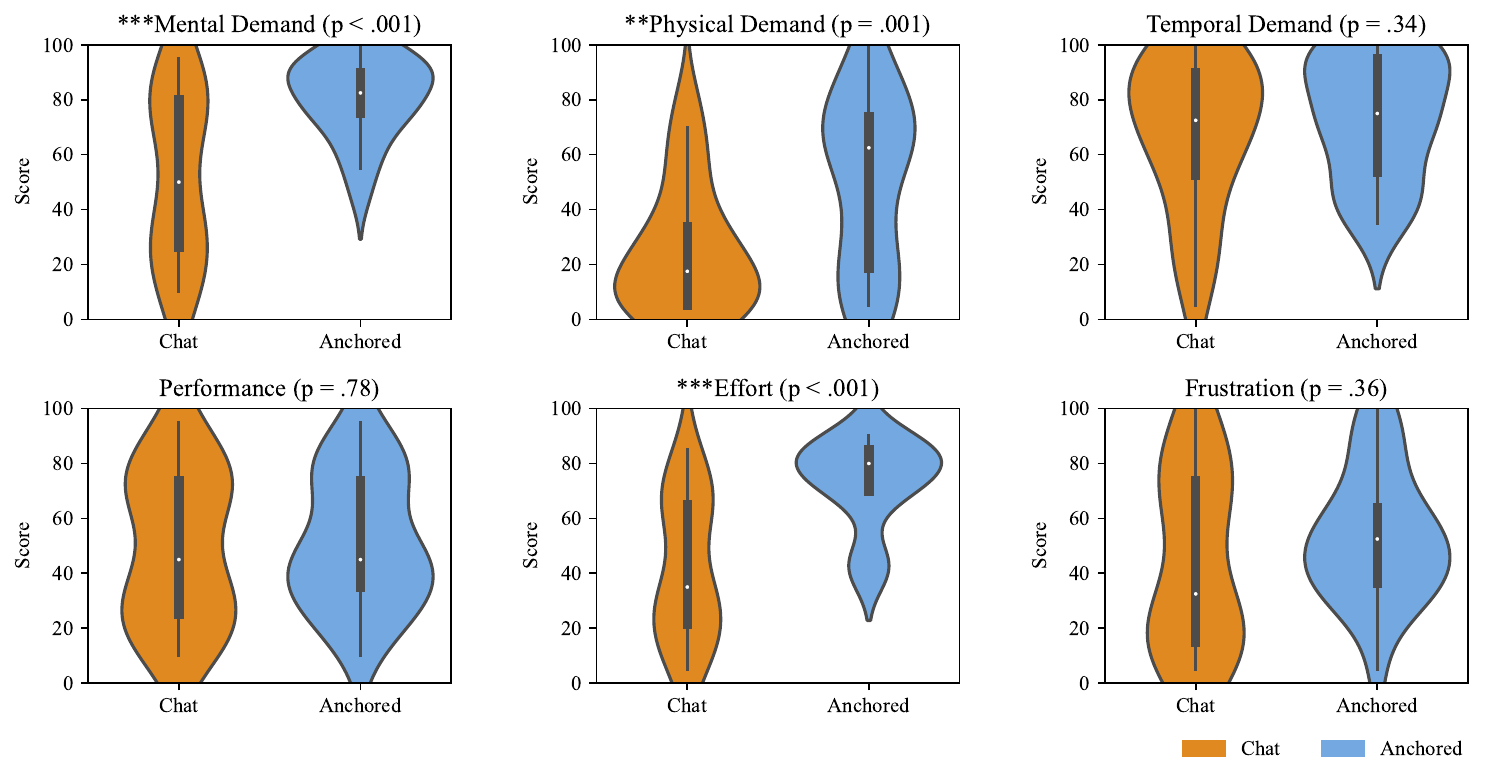}
    \caption{NASA-TLX workload ratings comparing Chat (orange) and Anchored (blue) conditions. Black bars represent interquartile ranges, white dots mark medians.}
    \Description{The violin plots compare subjective workload ratings between Chat (orange) and Anchored (blue) conditions across six NASA-TLX dimensions. Anchored produced significantly higher ratings of mental demand (p < .001), physical demand (p = .001), and effort (p < .001) compared to Chat. No significant differences were observed for temporal demand (p = .34), performance (p = .78), or frustration (p = .36). These results suggest that although Anchored increased cognitive and physical effort, it did not negatively impact participants' perceived performance or frustration levels.}
    \label{fig:tlx-result-plot}
\end{figure}

Our mixed-methods analysis reveals distinct differences in how users engage with AI feedback when presented through anchored comments vs. chat interfaces. We found that anchored comments promote more targeted, fine-grained revisions, whereas chat-based feedback leads to larger, wholesale text replacements (see Figure~\ref{fig:revision-plot}). This behavioral distinction is mirrored in writers' subjective experiences: the anchored AI comments significantly enhances their perceived ownership and control over the task process, indicating heightened sense of agency. Yet, these benefits are accompanied by an increase in perceived cognitive and physical effort, a trade-off participants oft-framed as necessary for deliberate engagement (see Figures~\ref{fig:survey-result-plot} and \ref{fig:tlx-result-plot}). Also, participants' qualitative reporting  suggests that the anchored format aligns more closely with writers' established mental models of human collaboration, likening the AI's role to that of a peer reviewer.

\subsubsection{\anchoredai{} promotes targeted, fine-grained revisions over wholesale replacement}
\label{revision_count_findings}

We examined how interface design influenced revision behaviors using a multivariate analysis of variance (MANOVA). Results showed a significant multivariate effect of interface (Chat vs. Anchored) on revision activity (Wilks' $\lambda = 0.73, F(3, 82) = 7.89, p < .001$). In contrast, task framing (Local vs. Global) did not yield a significant effect ($\lambda = 0.95, F(3, 82) = 0.47, p = .71$).

To better understand these differences, we conducted paired t-tests comparing the Chat and Anchored interfaces under both Local and Global conditions. Under the Local framing, participants using the Chat interface pasted significantly more words per paste ($M = 199.1 , SD = 113.0$) than those using Anchored ($M = 35.2 , SD = 59.8$), $t(21) = 7.29, d = 1.55, p < .001$. Similarly, the percentage of the document changed was higher with Chat ($M = 22.5 , SD = 18.5$) than with Anchored ($M = 6.8 , SD = 4.6$), $t(21) = 4.06, d = 0.87, p = .001$. However, the number of copy-paste action did not differ significantly between interfaces, $t(21) = 2.03, d = 0.43, p = .055$.

Under Global framing, the Chat interface again led to significantly more words pasted per action ($M = 185.8 , SD = 114.2$) compared to Anchored ($M = 68.9 , SD = 97.8$), $t(21) = 4.60, d = 0.98, p < .001$. No significant differences were found in the percent of document changed, $t(21) = 1.00, d = 0.21, p = .33$, or in the number of copy-paste actions, $t(21) = 1.24, d = 0.26, p = .23$.

Findings suggest that the Anchored interface fundamentally changes how participants approach revision. While the number of copy–paste actions remained similar across conditions—likely influenced by the time constraints of the task—the Anchored interface resulted in significantly fewer words being pasted overall. Under local revision tasks, we observed that the Chat interface can often produced a `splatter' effect, where changes extended into surrounding areas that did not necessarily require modification. This led to a greater proportion of the document being altered compared to the Anchored interface. In contrast, anchored feedback constrained edits to the relevant local context, producing more focused revisions. For global revision tasks, however, this distinction disappeared: because the goal was to address broader structural issues, both Chat and Anchored interfaces produced similar proportions of document change.

Qualitative data from follow-up interviews supported these quantitative results. Participants described using the Chat interface to make large-scale edits. For example, P20 said, \emph{``For the Chat interface, I kind of just copy-pasted the whole revised essay. I would read it over, like, yeah that looks good to me.''} In contrast, the Anchored interface prompted more deliberate engagement. As P20 continued, \emph{For the Anchored interface, because it gave suggestions a little bit at a time, it felt like I had to actually think about changes I wanted to apply.''}

Other participants shared similar experiences. P2 said, \emph{``I knew what it was looking at specifically and where [the Anchored interface] wants me to make the changes, whereas in [the Chat interface], it just did a mass edit and I didn't know what was going on.''} Similarly, P10 said, \emph{``It was nice that [the Anchored interface] was asking me about each change. So it was like... I had to think, 'Do I actually want this change in there?'''}

\subsubsection{\anchoredai{} enhances perceived control, ownership, and authorship.}
\label{survey_control_author_ownership}

\begin{figure}
    \centering
    \includegraphics[trim=0 0 0 0,clip,width=1.0\columnwidth]{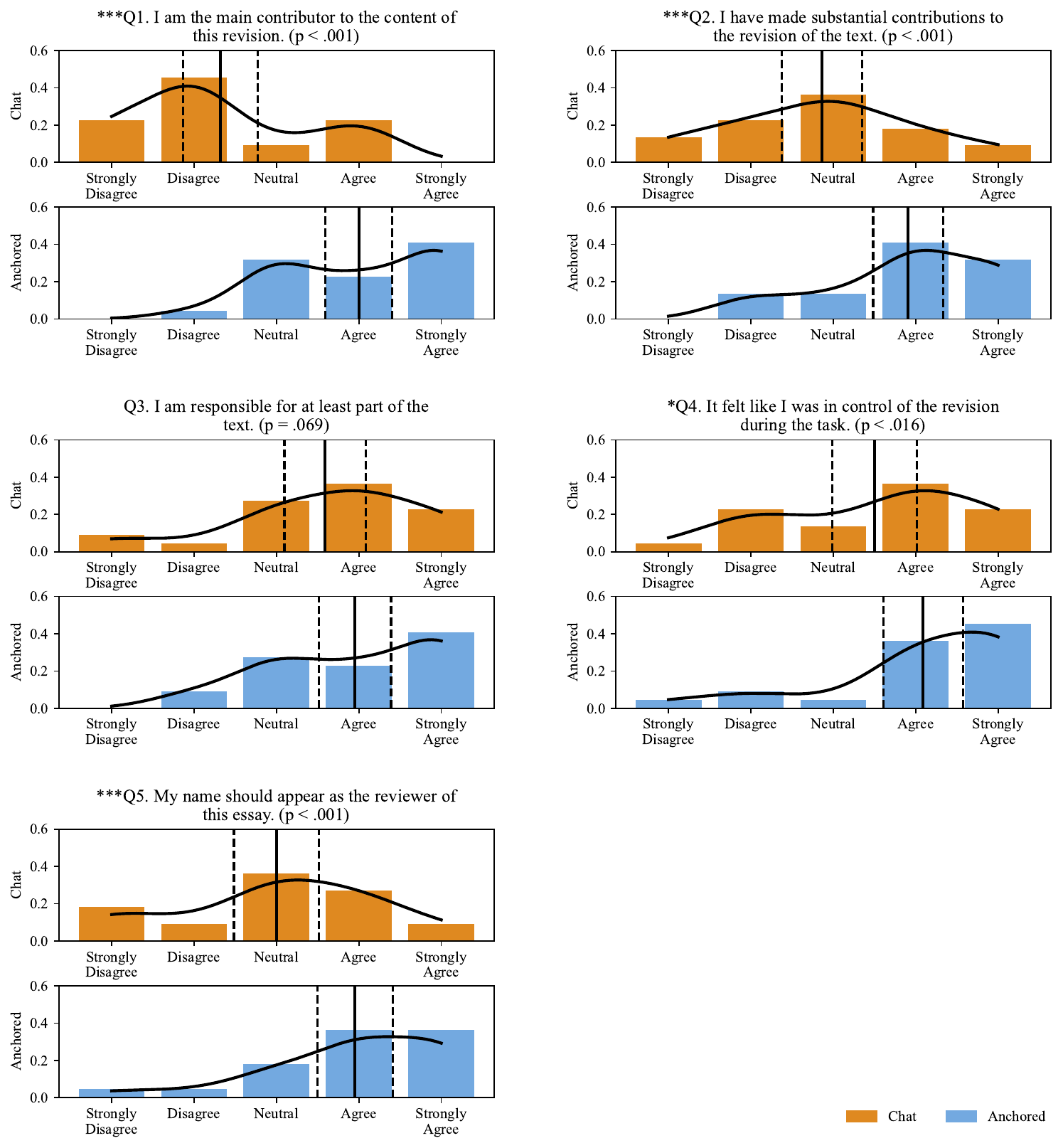}
    \caption{Response distributions for five Likert-scale questions assessing perceived control, ownership, and authorship across Chat (orange) and Anchored (blue) interfaces. Bars represent the proportion of responses, black curves show kernel density estimates, and vertical lines indicate the mean (solid) with 95\% confidence intervals (dashed). \anchoredai{} elicited higher agreement on questions related to control (Q1 and Q2) and showed upward trends for ownership (Q3) and authorship (Q5).}
    \Description{Figure of Likert-scale response distributions as bar charts for five survey items comparing Chat (orange) and Anchored (blue) conditions. Anchored consistently shifted responses toward stronger agreement with authorship and control statements. Participants using Anchored were significantly more likely to agree that they were the main contributor to the revision (Q1, p < .001), made substantial contributions (Q2, p < .001), and should be recognized as the reviewer of the essay (Q5, p < .001). Anchored also increased agreement with feeling in control during the revision task (Q4, p < .016), with Q3 (responsibility for part of the text) trending in the same direction (p = .069). These findings suggest that anchoring increased participants' sense of ownership and agency over the revision process relative to the chat-based interface.}
    \label{fig:survey-result-plot}
\end{figure}

To assess participants' sense of agency during revision, we administered five Likert-scale survey questions targeting perceived control, ownership, and authorship (see Table~\ref{tab:survey-questions}). As shown in Figure~\ref{fig:survey-result-plot}, responses for the Anchored interface were higher than Chat across all measures.

A Wilcoxon signed-rank test revealed significant differences between interfaces for four of the five questions. Participants using Anchored reported stronger agreement with the statements ``I am the main contributor...'' (Q1: $M = 4.0, SD = 1.0$ vs. Chat $M = 2.3, SD = 1.1$; $W = 34.0, p < .001$), ``I have made substantial contributions...'' (Q2: $M = 3.9, SD = 1.0$ vs. Chat $M = 2.9, SD = 1.2$; $W = 60.0, p < .001$), ``It felt like I was in control...'' (Q4: $M = 4.1, SD = 1.1$ vs. Chat $M = 3.5, SD = 1.2$; $W = 183.0, p = .016$), and ``My name should appear as reviewer...'' (Q5: ($M = 4.0, SD = 1.1$ vs. Chat ($M = 3.0, SD = 1.2$; $W = 34.0, p < .001$). Question 3 (``I am responsible for at least part of the text'') showed a positive trend, but did not reach statistical significance ($M = 4.0, SD = 1.0$ vs Chat $M = 3.0, SD = 1.2$; $W = 88.0, p = .069$). Overall, these findings indicate that the Anchored interface foster a greater sense of authorship and control over the revision process.

Interview data reinforced these findings. Participants consistently described the Anchored interface as promoting a greater sense of agency. P15 said, \emph{``With the [Chat interface] I was kind of just copying whatever it wrote... with the anchored one, it was more of my voice and I felt more in control of the changes.''} Similarly, P19 said, \emph{``[The Chat interface] definitely saves time... but I would feel better submitting the work that has [anchored feedback] with it. It just seems more of a contribution from yourself, because it suggests what to do---it doesn't tell you what to write.''}

Other participants emphasized the cognitive engagement prompted by Anchored feedback. P11 said, \emph{``I liked [the Anchored interface]. I definitely felt more in control... I was engaging my brain more... having assistance through suggestions is a lot easier to keep my brain engaged, rather than just copy pasting.''} P2 said, \emph{``I also felt more confident [using the Anchored interface]... because this is just a small suggestion, they didn't change the entire thing. I can delete this line or keep it. It was more clear, and I was making that decision consciously.''}


\subsubsection{The benefits of \anchoredai{} comes at the cost of increased cognitive and physical effort}
\label{findings_nasa_tlx}

We also evaluated participant's self-accessed workload using the NASA-TLX scale. Participants rated their experience across six dimensions: mental demand, physical demand, temporal demand, performance, effort, and frustration. Paired t-tests revealed that the Anchored interface was associated with significantly higher mental demand ($M = 80.7, SD = 14.7$ vs Chat $M = 51.8, SD = 30.3$; $t(21) = -4.74, d= -1.01 p < .001$), physical demand ($M = 50.7, SD = 31.0)$ vs. Chat ($M = 25.5, SD = 23.1$; $t(21) = -3.73, d=-0.80, p = .001$), and effort ($M = 73.0, SD = 15.9)$ vs. Chat $M = 41.4, SD = 25.7$; $t(21) = -5.20, d = -1.11, p < .001$). No significant difference were found for temporal demand ($t(21) = -0.98, d = -0.21, p = .34$), performance ($t(21) = -0.28, d = -0.06, p = .78$), or frustration ($t(21) = -0.93, d = -0.20 p = .36$).

Interview data provided further insight into these findings. Several participants acknowledged the increased mental effort required by the Anchored interface, but framed it as a positive aspect of the experience. P7 said, \emph{``[The Anchored interface] feels like something where you can actually critically think and be like, `How can I use this as a tool to express my own thinking better?'''} Similarly, P3 said \emph{``Definitely the anchored AI... a lot more thought is put into what has to be changed and what should be changed.''}



Several participants also described \anchoredai{} as closely resembling the experience of receiving feedback from classmates. For example, P16 said, \emph{``If I didn’t use AI, I would ask a peer for review... like direct comments, similar to how the \anchoredai{} does it.''} Others participants also emphasized the collaborative nature of the interaction. P7 added, \emph{``Normally I get friends to help me edit... I would say more so the anchor-based AI replicated that process.''} P8 said, \emph{``[The Anchored interface] is just like you ask somebody else to review your paper.''}

\section{Discussions}

Our findings reveal a trade-off: while chat-based interfaces offer a more direct path to a large-scale revision, anchored commenting promotes a more deliberate, targeted, and empowering editing process. We interpret these findings, situate them within prior work in HCI and CSCW, and discuss their broader implications.

\subsection{Anchoring as a Scaffold for Agency}

Our central finding is that the interface paradigm directly shapes the nature of human-AI collaboration in writing. The chat interface encouraged \emph{wholesale replacement}, where participants frequently copied and pasted large, AI-generated passages, leading to faster but less engaged revision. In stark contrast, \anchoredai{} fostered \emph{targeted, deliberate revision}. By localizing feedback to specific text spans, it prompted users to make smaller, more considered changes.

This behavioral distinction directly corresponds to the psychological outcomes we measured (see Section \ref{survey_control_author_ownership}). Supporting H2, \anchoredai{} significantly enhanced writer's perceived \emph{ownership, control, and authorship}. We argue that the spatial and semantic coupling of anchored comments acts as a \emph{cognitive scaffold}. It breaks down the complex, abstract task of ``document revision'' into a series of discrete, manageable, and contextualized decisions. This interrupts the path of least resistance---uncritical copy-pasting, creating a natural ``pause'' for evaluation at each anchor point. This aligns with research on cognitive forcing functions, which demonstrates how structured interaction design can nudge users toward more reflective engagement reduce overreliance on automated systems \cite{Bucinca-2021}. Our finding also share similar observation as Draxler et al.'s findings where a simple act of signing their name made users feel a stronger sense of ownership on content written in support of AI's generation \cite{Draxler-2024}. Overall, this helps mitigate the risk of the AI's role shifting from a supportive assistant to a directive author \cite{kosmyna-2025}.

\subsection{The Desirable Difficult of Anchored AI Feedback}

A key trade-off revealed in our study was that the benefits of \anchoredai{} came at the cost of increased effort (See Section \ref{findings_nasa_tlx}). While this might typically viewed as a usability flaw, we interpret it as a form of ``desirable difficulty,'' a concept from cognitive psychology where introducing certain challenges can lead to deeper processing \cite{bjork2011making}. The increased effort reported by our participants was not a symptom of a confusing interface, but rather a byproduct of meaningful engagement.  Participants' own words support this interpretation; they describe ``engaging [their] brain more'' and feeling that their own ``voice'' came through more strongly when using the anchored interface. The chat interface, by contrast, exacerbates the risks of automation bias and uncritical acceptance of AI suggestions \cite{bansal2021}. We argue that the moderate increase in effort is thus a valuable friction, ensuring that the writer remains an active participant in the creative process.

\subsection{Aligning AI with Human-Human Collaboration Models}

One of the most prominent qualitative findings was that participants spontaneously and consistently compared their experience with \anchoredai{} to receiving feedback from a human peer. They likened its functionality to the familiar process of a collaborator leaving comments in Google Docs or Microsoft Word. This suggests that the anchored paradigm aligns with writer's established mental models of collaborative writing, built around the principles of contextualized, asynchronous feedback. In contrast, the chat interface imposes a conversational model that is less suited to the contextualized task of targeted text revision. Our findings demonstrate that foundational principles from HCI and CSCW, such as the importance of comment-referent coupling for reducing ambiguity \cite{Marshall-1998, ohara-2002} and supporting awareness among collaborators \cite{birnholtz2012tracking}, are not just relevant but essential for effective human-AI collaboration.

This principle of transferring CSCW paradigms to human-AI interaction extends far beyond anchored comments. Consider the role of \emph{awareness mechanisms}, which are critical for coordinating collaborative work \cite{dourish1992awareness}. Future systems could provide users with a "revision history" for the AI's own suggestions, akin to DocuViz \cite{wang2015docuviz}, making its reasoning process more transparent and less of a "black box." Similarly, \emph{coordination protocols} could be adapted to manage the partnership. A writer might explicitly ``assign'' the AI the role of a proofreader, limiting its contributions to grammatical fixes, or the role of a brainstorming partner, encouraging more divergent suggestions. By explicitly defining roles and making the AI's actions and intentions more legible, these established CSCW patterns could transform the interaction from a simple command-response cycle into a more predictable, controllable, and ultimately more trustworthy collaboration, further enhancing the writer's sense of agency.

\subsection{Implications for Design}

These findings highlight an important design tension between enhancing user agency and increasing cognitive workload. Anchored AI clearly fostered a stronger sense of control, authorship, and ownership, yet it did so at the cost of higher mental demand and effort. This mirrors Tatar's observation that interaction design often involves tensional relationships \cite{tatar-2007}, where gains in one dimension (e.g., user agency) are accompanied by trade-offs in another (e.g., efficiency or ease). Rather than treating this as a limitation, such tensions can be understood as design resources: they clarify the compromises that must be navigated when tailoring systems to different goals and contexts. From this perspective, we out four key implications:

\begin{itemize}
    \item \textbf{Design for Deliberate Engagement, Not Just Frictionless Automation.} Anchored AI slowed participants down in productive ways, prompting them to evaluate suggestions more carefully and integrate changes selectively. This deliberate mode of revision reinforced writers' sense of authorship and accountability, making them feel entitled to claim the text as their own. Such effects mirror findings by Draxler et al. \cite{Draxler-2024}, who observed that even minimal acts of contribution—such as signing one’s name—can enhance perceived ownership. For educational and professional contexts where reflection and authorship are central, introducing this kind of ``productive friction'' may help cultivate deeper engagement with feedback, even if it requires greater effort.
    
    \item \textbf{Consider Task-Context Alignment.} The contrast between local and global revision tasks highlights that the benefits of anchoring are context-dependent. For local edits, anchoring focused revisions and reduced unnecessary spillover into surrounding text. For global edits, however, chat and anchored systems produced similar outcomes. This is similar to Tatar's observation that design often involves tensional relationships \cite{tatar-2007}, where improvements in one dimension (e.g., precision) may trade off against others (e.g., efficiency). Rather than seeking a universal `best' interface, designers should align the modality to the task: anchoring for precision and careful judgment, chat for breadth, speed, or brainstorming.
    
    \item \textbf{Leverage Familiar Collaborative Metaphors.} AI writing assistants should be designed to feel like collaborators, not black-box generators. Using established UI patterns from collaborative editors (e.g., comment threads, tracked changes, suggestions) can help set appropriate expectations for the AI's role and foster a more partnership-oriented dynamic between the writer and the system.
    
    \item \textbf{Explore Hybrid and Adaptive Workflows.} Participants' experiences suggest that future writing tools should not force a binary choice between anchored and chat modalities. Instead, systems could support fluid transitions between the two, or adaptively switch modes based on task demands. Prior work on shaping AI collaboration \cite{dhillon2024shaping} highlights the value of adaptable systems that give users control over how they engage with AI. A hybrid approach would allow writers to leverage the strengths of both modalities—anchoring for focused uptake of suggestions and chat for open-ended exploration—while mitigating their respective limitations.
\end{itemize}

\subsection{Limitations}

Our study has several limitations that constrain the generalizability of the findings. First, the participant pool consisted exclusively of university students. While this group represents an important population of writers, the results may not fully extend to professional or non-academic contexts. Second, we examined only one type of document and revision task. Writing and revision behaviors can vary substantially across genres (e.g., academic essays, creative writing, workplace reports), so additional studies are needed to evaluate \anchoredai{} in diverse settings. Furthermore, our evaluation focused on a meta-commentary revision task, which may emphasize certain aspects of feedback integration; results might differ in scenarios where writers are generating or substantially restructuring content.

The design of our prototype also shaped participants' experiences. Minor implementation details may have slightly dampened preferences for \anchoredai{} and constrained how smoothly participants interacted with the system, such as a slow loading experience for feedback and text based result loading. Because the tasks were assigned, participants may have been less invested in the outcomes than they would have been with their own authentic writing projects, although using standardized texts allowed us to better control for confounding factors. Finally, while our findings highlight clear benefits of anchoring, future work should also investigate conditions where \anchoredai{} may be less effective, or where differences between anchored and chat-based feedback diminish.

\subsection{Future Work}

Our findings point to several promising directions for advancing AI-supported writing interfaces. A key opportunity lies in developing systems that can dynamically decide when to anchor feedback and when to present it in a chat format, depending on the context and nature of the suggestion. Current designs treat anchoring and chat as mutually exclusive, but an adaptive system could flexibly switch between them or even combine the two—for example, anchoring to text when precision is needed, or offering meta-commentary in chat when broader guidance is more appropriate. Future work should also explore optimal levels of anchor specificity, as our prototype often defaulted to very fine-grained spans that were not always ideal; intelligent selection of sentence- or paragraph-level anchors may better balance precision and usability.

Beyond writing tasks, anchoring opens up possibilities across other domains. Applications in coding, multimedia authoring, and collaborative work could leverage anchoring to tie AI feedback directly to relevant resources, code fragments, or sections of a video. This resonates with Price's (1998) on linking across documents \cite{Price-1998}, suggesting opportunities to design AI comments that connect across multiple information sources—for instance, an anchored comment that also points to a relevant dataset. Additional directions include building typologies of semantic comment types, enabling users to filter or prioritize feedback by its role (e.g., stylistic vs. structural). Finally, given the strong transfer effect of the chat metaphor and participants' expressed desire for flexibility, future systems should explore hybrid interfaces that allow seamless movement between anchored and chat modes, and even support anchoring comments to multiple locations simultaneously. Such directions could extend the benefits of anchoring while retaining the familiarity and breadth of chat-based interaction.

\section{Conclusion}

\anchoredai{} demonstrates that the form of AI feedback matters as much as its content. By binding model suggestions to precise spans of text and maintaining those bindings through document change, our system re-centers the writer in the revision loop. Across a controlled comparison with a chat-based interface, we observed a characteristic shift in behavior: participants made more selective, localized edits, pasted fewer words, and changed a smaller share of the document, while reporting stronger feelings of control, authorship, and ownership. Revision quality remained comparable across interfaces. Together, these results suggest that anchoring can mitigate tendencies toward overreliance without sacrificing outcomes, reframing the AI from surrogate author to situated collaborator.

At the same time, our findings surface a productive tension: agency gains came with higher reported effort. Rather than treating this as a flaw, we view it as an actionable design trade-off that system builders can tune to context—privileging deliberate engagement in educational and professional revision, or speed and exploration during early ideation. Our technical contributions—optimally unique Anchoring Context Windows and update-aware context retrieval—offer a practical substrate for making such trade-offs explicit and controllable in future tools.

More broadly, the anchoring paradigm extends beyond prose. Tying AI feedback to concrete referents—lines of code, cells in a spreadsheet, regions in a video timeline—offers a general recipe for grounding assistance in collaborative production work. We envision hybrid interfaces that fluidly combine anchored and chat modalities, adaptively choosing when precision or breadth best serves the writer's goal. By designing for where AI speaks, not only what it says, we can build assistance that strengthens human judgment, preserves authorship, and supports intentional, high-quality revision.

\bibliographystyle{plainnat}
\bibliography{main}

\end{document}